\documentclass[useAMS]{aa}

\usepackage[latin1]{inputenc}
\usepackage[T1]{fontenc}
\usepackage{graphicx}
\usepackage{psfig}
\usepackage{natbib}
\usepackage{rotating}
\usepackage{amssymb}
\bibpunct{(}{)}{;}{a}{}{,} 

\def\gs{\mathrel{\raise0.35ex\hbox{$\scriptstyle >$}\kern-0.6em
\lower0.40ex\hbox{{$\scriptstyle \sim$}}}}
\def\ls{\mathrel{\raise0.35ex\hbox{$\scriptstyle <$}\kern-0.6em
\lower0.40ex\hbox{{$\scriptstyle \sim$}}}}

\def\kms{km\,s$^{-1}$}
\def\deg{$^\circ$}

\def\smm1{SMM\,J16359+6612}

\title{Molecular Gas in a $z\sim 2.5$ Triply-Imaged,\\ sub-mJy
Submillimetre Galaxy Typical \\of the Cosmic Far-Infrared Background
\thanks{Based on observations carried out with the
IRAM Plateau de Bure Interferometer. IRAM is supported by INSU/CNRS
(France), MPG (Germany) and IGN (Spain).}}
\titlerunning{IRAM Plateau de Bure Observations of \smm1}
\authorrunning{Kneib et al.}

\author{
Jean-Paul Kneib,\inst{1,2,3}
Roberto Neri,\inst{4}
Ian Smail,\inst{5}
Andrew Blain,\inst{2}\\
Kartik Sheth,\inst{2}
Paul van der Werf\inst{6} 
\and
Kirsten K.\ Knudsen\inst{6}
}
\institute{Observatoire Midi-Pyr\'en\'ees, CNRS-UMR5572, 
14 Avenue E.\ Belin, 31400 Toulouse, France
\and Caltech-Astronomy, MC105-24, Pasadena, CA 91125, USA
\and OAMP, Laboratoire d'Astrophysique de Marseille,
traverse du Siphon, 13012 Marseille, France
\and IRAM, 300 rue de la Piscine, 38640 Saint Martin d'H\`eres, France
\and Institute for Computational Cosmology, University of Durham,
South Road, Durham DH1 3LE, UK
\and Leiden Observatory, P.O.\ Box 9513, NL -- 2300 RA Leiden, The Netherlands
}

\date{Received ---, accepted ---.}

\begin{document}

\abstract{We present the results of observations from the Plateau de Bure
IRAM interferometric array of
the submillimetre (submm) galaxy \smm1 lying at $z=2.516$ behind the
core of the massive cluster A\,2218.  The foreground gravitational lens
produces three images with a total magnification of 45 of this faint
submm galaxy, which has an intrinsic submm flux of just $f_{850\mu{\rm
m}}=0.8$\,mJy -- placing it below the confusion limit of blank-field
surveys. The substantial magnification provides a rare opportunity to
probe the nature of a distant sub-mJy submm-selected galaxy, part of
the population which produces the bulk of the cosmic far-infrared
background at submm wavelengths. Our observations detect the CO(3-2)
line in all three images, as well as the CO(7-6) line and the dust
continuum at 1.3mm for the brightest image
but only at a 3$\sigma$ level.  The velocity profile of
the CO(3-2) line displays a double-peak profile which is well fit by
two Gaussians with FWHM of 220\,\kms\ and separated by 280\,\kms.  We
estimate the dynamical mass of the system to be $\sim 1.5\times
10^{10}$\,M$_\odot$ and an H$_2$ gas mass of $4.5\times
10^9$\,M$_\odot$.  We
identify a spatial offset of $\sim 1''$ between the two CO(3-2) velocity
components, again benefiting from the magnification due
to the foreground lens, modeling of which indicates that the offset
corresponds to just $\sim 3$\,kpc in projection at $z=2.516$.  The spatial
and velocity properties of these two components are closely related to
features detected in previously published H$\alpha$ spectroscopy.  We
propose that this source is a compact merger of
two typical Lyman-break galaxies with a maximal separation between 
the two nuclei of about 3\,kpc, although a dusty disk explanation is not 
excluded. This system is much less luminous and massive than
other high-redshift submillimetre galaxies studied to date, but it
bears a close similarity to similarly luminous, dusty starburst
resulting from lower-mass mergers in the local Universe.
 }

\maketitle

\section{Introduction}

%
%
\begin{figure*}
\centerline{\psfig{file=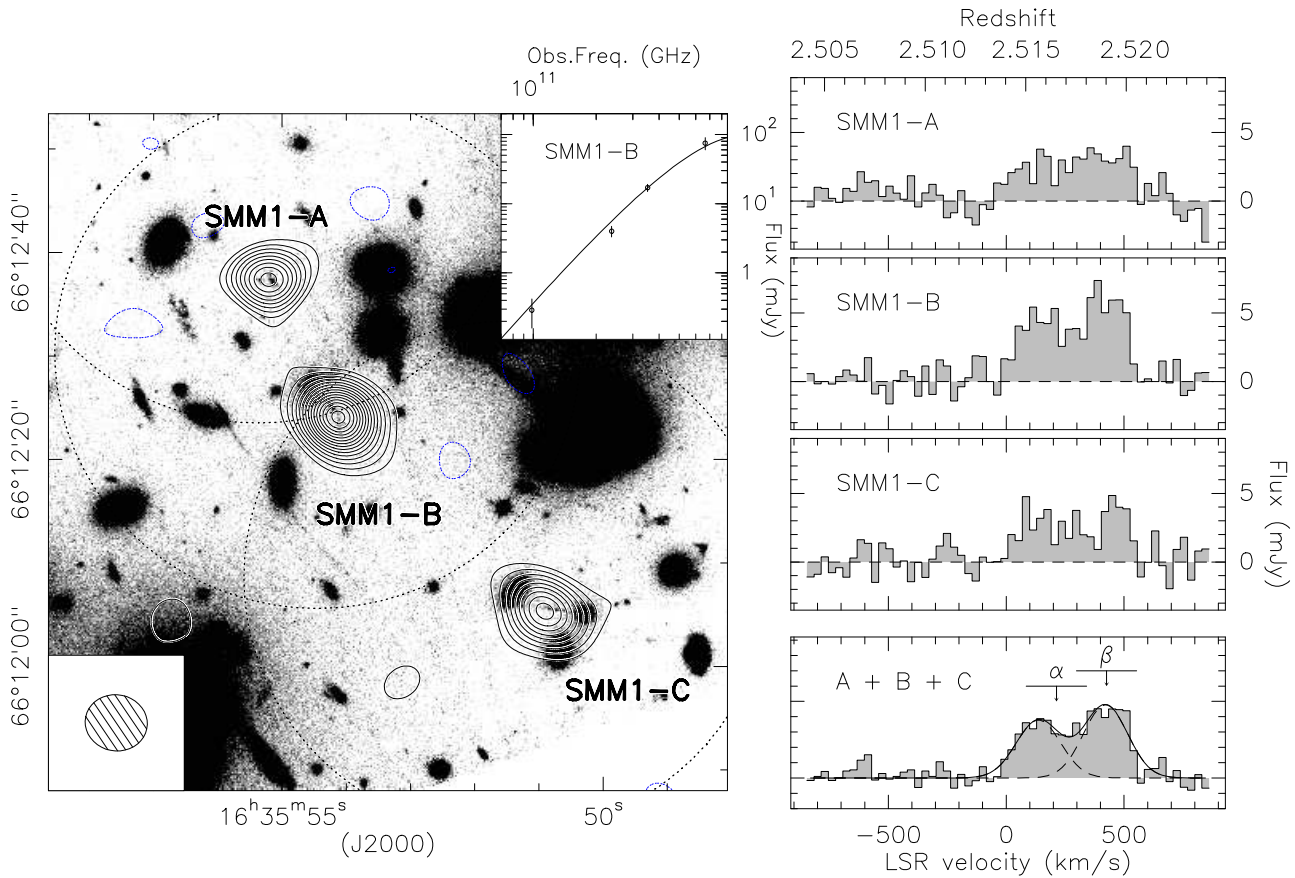,angle=-90,width=17cm}}
\caption{({\it Left}) The IRAM/PdBI $^{12}$CO(3-2) map of \smm1 (=SMM1)
superposed on the {\it HST} F814W image of A\,2218. We clearly identify
the images SMM1-A, B and C, all three well centered on their optical
counterparts. The contours start at $\pm2 \sigma$ and are plotted in
steps of $\sigma=0.24$\,mJy. The large dashed circle give the
half-power size of the primary beam of the PdBI at 98\,GHz at the three
pointing positions. The synthesized beam is displayed in the lower-left
corner and has a FWHM of $5.9''\times5.4''$ and a position angle of
98\deg.  We stress that the effective beam size in the source plane is
roughly 20 times smaller along the shear direction (PA of $\sim 40^\circ$)
in the vicinity of SMM1-B. The inset at top
right shows the spectral energy distribution of SMM1-B from the 3\,mm,
1.3\,mm, 850$\mu$m and 450$\mu$m measurements (the 450- and 850-$\mu$m
fluxes density is measured from the SCUBA discovery maps in Kneib et
al.\ 2004a).
({\it Right}) This panel displays (from top to bottom) the velocity
profile of the three different submm images SMM1-A, B and C, and of the
sum of these three spectra.  A double-peak profile is clearly observed
for SMM1-B and -C, as well as in the sum of all three components
(bottom panel). Fitting two Gaussians to the the sum of the three
velocity profiles indicates that the two components are separated by
$\sim$280\,\kms.  There is weak evidence that the higher-velocity
component has a higher flux density, but a narrower width.  For
comparison, we mark on the redshifts of the two features identified by
Kneib et al.\ (2004a) from their analysis of the {\it HST} imaging and
H$\alpha$ spectroscopy of this system: $\alpha$ and $\beta$ (see Figure
2).
}
\end{figure*}

The intensity of the far-infrared background indicates that
far-infrared luminous, dusty sources make a significant contribution to
the overall energy output from galaxies over the lifetime of the
Universe (Puget et al.\ 1996, Fixsen et al.\ 1998).  This emission may either represent AGN
activity, in which case these dusty sources must coincide with an
important phase in the growth of super-massive black holes, or
dust-obscured star formation, in which case they are responsible for
much of the massive star formation and metal enrichment occurring in
galaxies at high redshifts.  To understand the nature and properties of
this activity we need to resolve the background into individual
sources, suitable for more detailed study.  For sources contributing to
the peak of the far-infrared background, 100--200$\mu$m, this work
relies on space-based observatories, with modest apertures and
correspondingly bright confusion limits, resulting in the resolution
and identification of only a small (and hence perhaps unrepresentative)
fraction of the galaxies contributing to the background, $\ls 10$\%
(Dole et al.\ 2001).

Atmospheric windows at longer wavelengths allow larger-aperture
ground-based telescopes to resolve a greater fraction of the cosmic
background at millimetre (mm) and submillimetre (submm) wavelengths.
Continuum surveys in the submm/mm wavebands using both SCUBA on the
JCMT and MAMBO on the IRAM 30-m have identified several hundred of
submm/mm galaxies (hereafter SMGs) to date.  However, again the relatively
coarse resolution of these telescopes, $\gs 10$\arcsec, results in
source confusion limiting these observations to identifying individual
sources brighter than $\gs 2$\,mJy at 850\,$\mu$m (e.g.\ Blain et al.\
2002).  Thus most of the best-studied samples of SMGs have 850-$\mu$m
fluxes in the range 4--10\,mJy (e.g.\ Ivison et al.\ 2002; Webb et al.\
2003; Bertoldi et al.\ 2000), which have also been the focus of recent
work to determine the redshift distribution of the bright submm
population, yielding a median redshift of $z=2.3$ (Chapman et al.\
2003, 2004).  Unfortunately, once again, this population may not be completely
representative as they only contribute around 30\% of the far-infrared
background at submm wavelengths (Smail et al.\ 2002).

To access the submm sources which are responsible for the bulk of the
cosmic submillimetre background (and hence contain most of the
star-formation or AGN activity) we need to probe down to flux densities
below 1\,mJy at 850\,$\mu$m (Smail et al.\ 2002, Cowie et al.\ 2002).  
These sources are
fainter than the confusion limit of blank-field submm surveys making
them impossible to identify or study individually in detail from such
samples.  Nevertheless, they are amenable to study by
exploiting the natural gravitational magnification of massive clusters
of galaxies (Kneib et al.\ 1996; Blain 1997).  The magnification
provided by the cluster lens both boosts the brightness of the source
and provides a finer effective beam size -- reducing the effects of
confusion -- and bringing the sub-mJy submm population within reach of
current telescopes. This technique was used for the first
extragalactic submm survey (Smail et al.\ 1997; Blain et al.\ 1999;
Chapman et al.\ 2000; Cowie et al.\ 2002) and has provided a powerful
route to understand the nature of this population of distant, faint
galaxies (Blain et al.\ 2002) as well as other high-redshift galaxy
populations (e.g.\ Baker et al.\ 2004).  These studies show that SMGs
with 850-$\mu$m fluxes within a factor of two of 1\,mJy have a surface
density of $\sim 3$--4 arcmin$^{-2}$ and produce roughly half of the
submm background (Smail et al.\ 2002).  Combining these SMGs with the
population of brighter sources accessible in blank-field surveys, we
can trace the properties of SMGs that contribute the majority of the
background radiation at submm wavelengths.

The first redshifts of SMGs were obtained for sources lensed by massive
galaxy clusters (Ivison et al.\ 1998, 2000) and cluster lenses continue to
aid in the spectroscopic identification of this population (e.g.\
Frayer et al.\ 2003; Kneib et al.\ 2004a; Borys et al.\ 2004).  These
redshifts enable us to search for the corresponding line emission from
cold gas within these galaxies -- providing valuable information on the
dynamics of these systems, the role played in these by the baryonic
component and the extent of the gas reservoir available to power future
activity.  The first such observations targeted the rotational
transitions of $^{12}$CO gas with the OVRO (Frayer et al.\ 1998, 1999,
Ivison et al.\ 2001)
and IRAM arrays (Downes \& Solomon 2003; Neri et al.\ 2003; Genzel et
al.\ 2003).  Similar work on brighter blank-field SMGs has built upon
the recent SMG redshift survey of Chapman et al.\ (2003, 2004), in a
major programme to detect CO line emission from a large sample of
bright submm galaxies with the IRAM PdBI (Neri et al.\ 2003; Greve et
al.\ 2005).  However, all of the submm galaxies studied in CO so far
(both cluster lens and blank-field targets) are intrinsically luminous
systems, none are from the mJy-submm population which dominates the
far-infrared background.

Recently, Kneib et al.\ (2004a) and Borys et al.\ (2004) have discovered
two strongly magnified and multiply-imaged SMGs at $z=2.516$ and
$z=2.911$ lensed respectively by the massive clusters A\,2218 and
MS\,0451$-$05.  These sources have intrinsic submm fluxes below 1\,mJy,
but with magnifications greater than 10 for at least two of the three
detected images in each system. Their CO emission should be within
reach of interferometric observations with current instruments.

In this paper, we present recent observations with the IRAM array of
redshifted molecular CO emission in the 3-mm and 1.3mm bands of the
triply-imaged sub-mJy SMG \smm1\ at $z=2.516$. Interferometric CO
observations of this system have recently been published by Sheth et
al.\ (2004) using the OVRO array, although the significance of their detections
is lower than that described here.  We present our observations in \S2,
describe our analysis and results in \S3 and give our conclusions in
\S4.  Throughout the paper we will assume an $\Omega=0.27$,
$\Lambda=0.73$ cosmology with $H_0=71$\,km\,s$^{-1}$\,Mpc$^{-1}$. At a
redshift of $z=2.516$ the angular scale is thus 8.18 kpc/arcsec.  All
quoted coordinates are in J2000.

\section{Observations}

%
%
\begin{table*}
\caption{Observed properties of the three images of SMM\,J16359+6612,
uncorrected for lens magnification.  We list the centroids of the
CO(3-2) emission and the flux-weighted redshifts derived for each
image.  $I_{\rm CO(3-2)}$ is the velocity-integrated line intensity
from the Gaussian fit (see text).  $I_{\rm B}/I_{\rm R}$ is the
corresponding ratio of the blue-shifted to red-shifted CO(3-2) line
emission.  $I_{\rm CO(7-6)}/I_{\rm CO(3-2)}$ is the intensity ratio of
CO(7-6) versus CO(3-2) line emission in the 0--370\,\kms\ window (with
respect to $z=2.514$, Figure~1).  $S_{\rm 1.3mm}$ is the flux density
of the 1.3mm continuum detected towards SMM1-B, the brightest
component. The last column gives the magnification factor at the
position of the three images derived from the detailed mass model of
the foreground cluster by Kneib et al.\ (2004a,b).
\label{tab.results}}
\begin{center}
\begin{tabular}{cccccccr}
ID & $\alpha$,$\delta$ & $z$ 
   & $I_{\rm CO(3-2)}$ & $I_{\rm B}/I_{\rm R}$ & $I_{\rm CO(7-6)}/I_{\rm CO(3-2)}$ & 
   $S_{\rm 1.3mm}$ & Mag \\
   & ($J2000$) &  & (Jy\,\kms)  &  &  &  (mJy)  &   \\
\hline
A & 16:35:55.18~66:12:37.2 ($\pm$0.2$''$) & 2.51740 
  & 1.67 ($\pm$0.13) & 0.74 ($\pm$0.02) & 2.8 ($\pm$1.2) & ... & 14 ($\pm$2) \\
B & 16:35:54.10~66:12:23.8 ($\pm$0.2$''$) & 2.51741 
  & 2.50 ($\pm$0.12) &  0.74 ($\pm$0.02) & 1.4 ($\pm$0.4) & 3.0 ($\pm$0.7) & 22 ($\pm$2) \\
C & 16:35:50.96~66:12:05.5 ($\pm$0.3$''$) & 2.51744 
  & 1.58 ($\pm$0.17) &  0.88 ($\pm$0.02) & ...         & ... &  9 ($\pm$2) \\
\hline
\end{tabular}
\end{center}
\end{table*}

%
%
\begin{figure*}
\centerline{\psfig{file=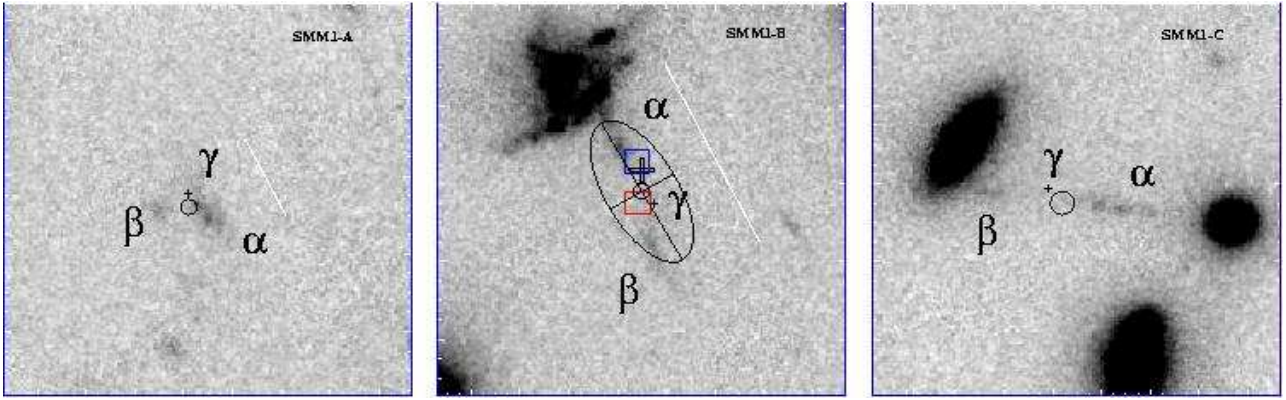,angle=0,width=17cm}}
\caption{10\arcsec$\times$10\arcsec\ regions from the F814W-band {\it
HST} WFPC2 image of the triply-imaged submm source \smm1, showing
images A, B and C (from left to right).  We overlay on each the
position of the CO(3-2) line emission centroid as measured from our
data (small circle), the relative uncertainty on these positions is
$\ls 0.3''$.  For component SMM1-B (middle panel) we plot the size of
the beam-corrected CO(3-2) profile as an ellipse and we also indicate
the centroid of the blue-shifted ($\le$220\kms) and red-shifted
($\ge$280\kms) velocity components of the line (open squares) and the 
location of the 1.3mm continuum emission (large open cross). Note that 
every image of SMM1 comprises a NIR source ($\gamma$, marked as a cross) 
which is bracketed by two bluer features visible in these F814W images 
($\alpha$ and $\beta$).  For SMM1-A and SMM1-B we have indicated by white 
lines the extent of the H$\alpha$ line observed in the NIRSPEC data of 
Kneib et al.\ (2004a).  The resolution of the {\it HST} images are
$\sim$0.15\arcsec; North is up and East is to the left. 
\label{fig.truecol}}
\end{figure*}

We used the six antennas of the IRAM PdBI to image the CO(3-2) and
CO(7-6) lines of \smm1 (named hereafter SMM1 in the text) at a redshift
of 2.516 (Kneib et al.\ 2004a). As the primary beam of the antenna is
about 50\arcsec\ and 21\arcsec\ at the frequency of redshifted CO(3-2)
and CO(7-6) (98.405\,GHz and 229.544\,GHz respectively), three
different positions around \smm1 were targeted to completely map the
three different images.  We have approximately 16\,hrs on-source at 3mm
and 5\,hrs at 1.3mm (after flagging low-quality visibilities) for the
central pointing at $\alpha$=16:35:54.5, $\delta$=66:12:31,
$\sim$9\,hrs at 3mm on the southern pointing ($\alpha$=16:35:51.7,
$\delta$=66:12:10) and $\sim 2$\,hrs on the northern one
($\alpha$=16:35:55, $\delta$=66:12:50).  No useful integration time was
obtained at 1.3mm at the latter two positions.  All these observations
were carried out in the D configuration between November 2003 and June
2004. To improve the 1.3mm continuum detection, we also used data from
an earlier $\sim$5\,hrs track which we observed on December 10, 2001 at
237.191\,GHz with the phase tracking center at $\alpha$=16:35:54.1,
$\delta$=66:12:23. We used 1637+574 as phase and amplitude calibrator,
and 0420$-$014, 0923+392, 3C84 and MWC349 as bandpass calibrators and
to define the flux density scale. We estimate that the flux densities
at 3mm are accurate to better than 10\%, and to better than 20\% at
1.3mm.

Our 3-mm spectra provide velocity coverage extending from $-$850 to
+850\,\kms\ around a zero-LSR redshift of $z=2.514$ for each of the
three sources.  3-mm line visibilities were averaged in frequency and
combined to produce natural-weighted dirty maps. These were
cleaned using the Clark algorithm, restored with a Gaussian beam of
5.9$''\times5.4$\arcsec\ with a position angle of 98\deg, and corrected
for primary beam attenuation. Natural-weighted dirty maps were also
made of the continuum emission at 3mm and 1.3mm by uv-averaging over
the line-free channels.  We estimate the absolute precision of the
astrometry of our maps as 0.2--0.3$''$, sufficient to precisely locate
the CO emission relative to sources in the {\it HST} imaging.  The
astrometry of the {\it HST} imaging is tied through wide-field
ground-based images to the USNO reference frame, yielding absolute
astrometry at the 0.2$''$ level (Kneib et al.\ 2004a).

\section{Analysis and Results}

%
%
\begin{table*}
\caption{The properties of the velocity components of the CO(3-2) and
H$\alpha$ lines of SMM1-B. The coordinate offsets are computed with
respect to the position of the optical feature $\alpha$ within SMM1-B
($\alpha$=16:35:54.19, $\delta$=66:12:24.92, Kneib et al.\ 2004a) --
note there the relative astrometry of the optical and CO maps is
uncertain at the $\sim 0.4''$ level.  CO(3-2) and H$\alpha$ velocities
are relative to the H$\alpha$ redshift of $\alpha$ of $z=2.5165$
(corresponding to $+210$\kms\ relative to $z=2.5140$).
\label{tab.line} 
}
\begin{center}
\begin{tabular}{crrcrr}
SMM1-B & \multicolumn{1}{c}{$\Delta\alpha$} & \multicolumn{1}{c}{$\Delta\delta$} & \multicolumn{1}{c}{$z$}  & \multicolumn{1}{c}{$v$} & \multicolumn{1}{c}{$\Delta v$}  \\
Component   & \multicolumn{1}{c}{(\arcsec)} & \multicolumn{1}{c}{(\arcsec)} &  & \multicolumn{1}{c}{\kms} &  \multicolumn{1}{c}{\kms} \\
\hline
CO(3-2) blue+red  & $-$0.56 ($\pm$0.16) &  $-$1.14  ($\pm$0.16)
        &  2.5172 ($\pm$0.0002) & 70 ($\pm$20) & 440 ($\pm$80) \\
\noalign{\smallskip}  
CO(3-2) blue & 0.42 ($\pm$0.20) & $-$0.37 ($\pm$0.20) & 2.5156 ($\pm$0.0002) &
	$-$70 ($\pm$20) & 220 ($\pm$50) \\
$\alpha$ ($H\alpha$) & 0.0 & 0.0 & 2.5165 ($\pm$0.0015) & 0 &
        280 ($\pm$60) \\ 
\noalign{\smallskip}
CO(3-2) red & 0.45 ($\pm$0.17) & $-$1.41 ($\pm$0.17) & 2.5189 ($\pm$0.0002) &
        210 ($\pm$20) & 220 ($\pm$50) \\
$\beta$ ($H\alpha$) & 0.78 ($\pm$0.10) & $-$2.51 ($\pm$0.10) &
        2.5190 ($\pm$0.0015) & 220 ($\pm$60) & 280 ($\pm$60) \\ \hline
\end{tabular}
\end{center}
\end{table*}

\subsection{The CO(3-2) line}

We detect the CO(3-2) line at high significance in all three images and
show the velocity-integrated line map and spectra in Figure~1, with
tabulated information in Table~1.  The flux-weighted mean redshifts of
the lines (integrated across the FWZI from $-10$ to 540\kms) in the
three images are identical -- $z_{\rm CO}=2.5174\pm 0.0002$.  The ratio
of CO(3-2) line intensities for the three images roughly corresponds to
that seen for their 850-$\mu$m and $K$-band fluxes from Kneib et al.\
(2004a): $I_{A/B}= 0.67\pm 0.18 $ and $I_{C/B}= 0.63\pm 0.19$, compared
to 0.64, 0.53 and 0.53, 0.40 from the 850-$\mu$m and $K$-band
respectively.  These flux ratios are all consistent with the predicted
magnification ratios from the lens model of the cluster:
$M_{A/B}=0.64\pm 0.14$ and $M_{C/B}=0.41\pm 0.12$.  The CO(3-2) line
widths measured from the spectrum of each image are also all large,
corresponding to FWZI of 500\,\kms.  These observations allow us for
the first time to categorically state that all three sources, SMM1-A, B
and C, represent images of a single background object.

Compared to the OVRO observations of \smm1\ in Sheth et al.\ (2004),
we measure a slightly lower total integrated CO flux: $5.75\pm
0.25$\,Jy\,\kms\ (versus $6.3\pm 0.2$\,Jy\,\kms), and a slightly higher
redshift: $z=2.5174\pm 0.0002$ (compared to $z=2.5168\pm 0.0003$).  In
both cases the differences are relatively minor and the higher signal
to noise of our detections would suggest our results are the more
reliable.  

We can also exploit the better signal-to-noise of our observations to
look in more detail at the structure in the CO(3-2) lines. We find that
the CO(3-2) line profiles display a characteristic double-peaked
profile in SMM1-B and SMM1-C, while the structure of the spectrum of
SMM1-A is less obviously bimodal.  We combine the three spectra to
increase the signal-to-noise yet further (assuming that differential
magnification effects across the source are small, which should be 
correct for a compact object) as shown in Figure~1. We fit the combined
velocity profile using a two Gaussian fit to determine the redshift,
width and relative intensity of the two velocity components. We find
that the velocity profile requires both Gaussians for an adequate fit,
each with widths of 220$\pm$20\,\kms\ and separated by
280$\pm$20\,\kms. The flux ratio between the blue-shifted and
red-shifted components is $I_{\rm B}/I_{\rm R}=0.74\pm0.02$ (Table~1).
The redshifts of the two velocity components are $z_{\rm
blue}=2.5156\pm0.0002$ and $z_{\rm red}=2.5189\pm0.0002$.  We note that
Kneib et al.\ (2004a) reported $z=2.5165\pm0.0015$ for the H$\alpha$
redshift of the optical structure $\alpha$, and $z=2.5190\pm0.0015$ for
feature $\beta$ in SMM1-B (Figure~2) and return to this point below.

To investigate these two velocity components in more detail and search
for changes in size or position with velocity, the visibilities were fitted
first with elliptical Gaussians. Although we could not resolve SMM1-A or -C,
SMM1-B (the most magnified image) is best fit by an elliptical Gaussian
with a beam-corrected size of $4.1\pm 0.6\arcsec\times
1.8\pm1.0$\arcsec and a position angle of $30\pm 13$\deg\ -- this is
plotted as an ellipse in the central panel of Figure~2. We note that
this agrees well with the size and orientation of the optical
counterpart of this system in Figure~2.  Correcting for the lensing
magnification, using the well-constrained mass model of A\,2218 (Kneib
et al.\ 1996, 2004b; Ellis et al.\ 2001), this ellipse maps to a linear
size of $\sim 3\times1.5$\,kpc in the source plane (there is some
complex, but weak, dependence according to the adopted position angle).
Note that the tentative resolution of SMM1-B claimed by Sheth et al.\
(2004) has a nearly orthogonal orientation to that derived here (PA of
$-68$\deg), transverse to the local shear direction and hence the
lensing-corrected size they estimate is significantly larger than our
measurement: 12\,kpc versus 3\,kpc.

To investigate whether the red-shifted and blue-shifted velocity
components originate from two different spatial positions, for the
blue-shifted emission we integrated the velocity map for all three
images from $\sim20$--230\,\kms\ (note the reference redshift is
$z=2.5140$) and for the red-shifted emission from $\sim
300$--510\,\kms. Again for SMM1-A and SMM1-C, no strong distinction
could be made in the position of the blue-shifted and red-shifted
velocity channels.  However, for the SMM1-B component the two velocity
components show a well-detected spatial separation (see Table~2 and
Figure~2) between the blue-shifted and red-shifted gas.  The two
components are separated along the major axes of the CO(3-2) emission
and the optical counterpart (Figure~2). Based on just the spatial
information, and allowing for the $\sim 0.4''$ uncertainty in the
relative astrometry of the CO and optical maps, we suggest that the
blue-shifted component of the SMM1-B CO(3-2) line is likely
associated with the optical feature $\alpha$ and the red-shifted 
component with the optical feature $\beta$.  
If this is indeed the case, then the blue- and
red-shifted components of the CO(3-2) emission in SMM1-B are both
spatially and kinematically coincident with the optical structures
$\alpha$ and $\beta$ cataloged by Kneib et al.\ (2004a).
Alternatively, the CO(3-2) line could be associated to a smaller rotating
structure closer and around component $\gamma$, however with no strong
optical or H$\alpha$ counterpart, which makes this possibility less likely
but not impossible.  Higher resolution millimeter observation of this system
should confirm the true nature of this system.

\subsection{The CO(7-6) line}

The velocity coverage of our parallel observations of the CO(7-6)
\smm1\ in the 1.3mm band extends from $-380$\,\kms\ to +380\,\kms\
(corresponding to the redshift interval $z=2.510$--2.518).  Thus, only
part (about 70\%) of the expected broad CO(7-6) line is covered by the
frequency setup. The noise level in the 1.3mm band is also relatively
high and does not allow a high confidence level measurement.  
Within this limited 1.3mm window we measure a CO(7-6) flux of
$I_{\rm CO(7-6)}=3.3\pm 1.4$ Jy.km/s for   SMM1-A and
$I_{\rm CO(7-6)}=2.5\pm 0.7$ Jy.km/s for   SMM1-B.
In order to quote representative measurements of the CO(7-6)/CO(3-2) line ratio,
we have computed the integrated flux in the redshift window from
$z=2.5140$ to $z=2.5185$ for the two lines CO(3-2) and CO(7-6) at the
position of SMM1-A and SMM1-B.  The derived flux ratios are: $I_{\rm
CO(7-6)}/I_{\rm CO(3-2)}=2.8\pm1.2$ for SMM1-A, and $I_{\rm
CO(7-6)}/I_{\rm CO(3-2)}=1.4\pm0.6$ for SMM1-B (the larger error for
SMM1-A is mainly due to the larger primary beam correction at 1.3mm for
this source). These measurements should be reduced by $\sim50$\% to
account for the contribution from continuum emission ($\sim3$\,mJy)
within the velocity window used to integrate the CO(7-6) line.

The mean CO(7-6) to CO(3-2) flux ratio for SMM1-A/B is comparable to
the value $I_{\rm CO(7-6)}/I_{\rm CO(3-2)} =1.14 \pm 0.20$ found for
the lensed SMG, SMM\,J14011+0252 at $z=2.565$ (Downes \& Solomon 2003).
As in SMM\,J14011+0252, the strength of
CO(7-6) relative to CO(3-2) indicates both a relatively high
temperature, $T\sim 50$\,K, and high H$_2$ density, $\sim
10^3$\,cm$^{-3}$, for the gas reservoir in this galaxy.

\subsection{Continuum emission}

We detected weak continuum emission towards SMM1-B at
$\alpha=$16:35:54.1, $\delta=$66:12:24.36.  The flux density is
$3.0\pm0.7$\,mJy at 1.3\,mm (237\,GHz), and within our astrometric
accuracy ($0.2''$--0.3\arcsec) this position (Figure 2) is coincident with the
integrated CO(3-2) emission from SMM1-B (the synthetised beam has a
size of $1.9''\times1.5''$ at a PA of $-37$\deg). We also detect a much
weaker continuum source at this precise position in our 3\,mm (98\,GHz)
data, with a flux density of $0.29\pm0.13$\,mJy.  We fit the spectral
energy distribution (SED) of the dust emission using these two flux
points, in combination with the 450- and 850-$\mu$m values from Kneib
et al.\ (2004a), see inset panel in Figure~1.  We determine a spectral
index of $\sim \nu^{3.2}$ from the slope of the SED between 850\,$\mu$m
and 3\,mm -- characteristic of optically thin dust emission.  The dust
emissivity index is thus $\beta\sim 1.2$, very similar to that derived
for bright {\it IRAS} galaxies locally ($\beta\sim 1.3\pm 0.2$, Dunne
et al.\ 2000).

We can also estimate the temperature from the fit to the four
long-wavelength photometric observations of the SMM1-B, this is
determined rather accurately, as $T_d=51 \pm 3$\,K. Including the 
Metcalfe et al.(2003)  ISO limit we have  $T_d=48 \pm 3$\,K for
SMM1-B data assuming $\beta=1.5$. We can use this
value, along with the dust emissivity index and the
magnification-corrected 850\,$\mu$m flux (which is our most precise
flux measurement) to estimate the dust mass, assuming a dust mass
opacity coefficient, $\kappa_d= 0.077$\,m$^2$\,Kg$^{-1}$ at 850\,$\mu$m
(Dunne et al.\ 2000).  This calculation yields a mass of $(1.9 \pm 0.3)
\times 10^{7}$\,M$_\odot$, subject to the considerable uncertainty in
the opacity coefficient.
 
The corresponding total bolometric luminosity is $1.6\pm 0.4 \times
10^{12}$\,L$_\odot$. Hence, although intrinsically faint at millimetre
and submillimetre wavelengths, owing to its high dust temperature,
\smm1\ is still a relatively luminous galaxy. It will be interesting to
include {\it Spitzer Space Telescope} data to confirm and better
constrain the dust temperature  but unfortunately there is currently no
deep radio image of A\,2218.

\section{Conclusions}

We have compared the results of our PdBI observations of \smm1\ with
those derived from shallower data taken with the OVRO array by Sheth et
al.\ (2004).  We find broad agreement between these datasets, some
discrepancies do exist but can be attributed to the lower signal to noise of
the OVRO data.  The high-fidelity CO maps provided by the IRAM PdBI 
(in particular due to its better coverage of all 3 sources) add
significantly more detail to address the nature and properties of this
unique source.

A detailed analysis of the CO(3-2) velocity profile shows a double-peak
structure with a significant spatial offset between the blue- and
red-shifted gas components.  Comparing the positions and redshifts of
these two gas components with the complex optical/near-infrared
morphology discussed by Kneib et al.\ (2004a), we tentatively
identify the kinematic
components seen in CO(3-2) with two features visible in the {\it HST}
imaging of this system: $\alpha$ and $\beta$ (Kneib et al.\ 2004a). We
note that the CO(3-2)/H$\alpha$ luminosity ratio is much larger for
component $\beta$, either due to stronger dust obscuration in $\beta$
or because of slit-losses in the H$\alpha$ observations resulting from
a misalignment of the slit across the source. Near-infrared integral
field spectroscopy of this SMG to measure the spatial structure of the
H$\alpha$ line will be essential to better determine the relative
distribution of cold and ionised gas across this system.

The spatial separation of the two CO(3-2) components is $\sim1.5$\,kpc
(with a maximal extent of $\sim 3$\,kpc) on the sky (once corrected
for the lensing magnification). This scale suggests we are either
seeing a molecular gas disk or ring within a single galaxy, or two
strongly interacting gas-rich components of a merger.
The latter possibility is favored, as the three optical/NIR component
$\alpha, \beta$ and $\gamma$ are not co-linear, and thus are not 
representative of a stable disk.
We can use the spatial and velocity information (a rotation velocity 
of 150\,\kms\ at a scale of 3\,kpc) to estimate the dynamical masses in 
these two scenarios.  We derive a mass of M$_{dyn}=(0.8\pm0.2)\times
10^{10}\,$M$_\odot$ at a radius of 1.5\,kpc if the gas is distributed
in a ring/disk, and M$_{dyn}=(1.5\pm0.3)\times 10^{10}\,$M$_\odot$
within the same aperture if the system is a merger
(we assume a 90$^\circ$ inclination here, otherwise
both of these estimates need to be scale with $(\sin i)^{-2}$).

\smm1\ is the third SMG in which spatially-resolved CO emission has
been detected, the other two being SMM\,J02399$-$0135 at $z=2.81$
(Genzel et al.\ 2003) and SMM\,J14011+0252 at $z=2.56$ (Downes \&
Solomon 2003; Ivison et al.\ 2001).  SMM\,J02399$-$0135 shows a
double-peaked CO(3-2) line with a spatial offset of 10\,kpc and
400\,\kms\ between the components, while SMM\,J14011+0252 has only a
single velocity component (FWHM of 200\,\kms) with a size of $\sim
3$\,kpc (adopting the magnification factor for the lensing
configuration for this source from Swinbank et al.\ 2004).  The gas
distribution in SMM\,J14011+0252 is thus very similar in scale (as well
as having similar dynamical and far-infrared luminosities) to \smm1,
while SMM\,J02399$-$0135 is both significantly more luminous, larger
and more massive.  However, the dense gas detected in all three systems
has a much wider distribution than the comparable component in local
Ultraluminous Infrared Galaxies (ULIRGs), $\sim 0.3$\,kpc (Downes \&
Solomon 1998).

With our assumed cosmology, the total CO(3-2) line flux converts to an
observed CO(3-2) luminosity of SMM1-B of
$(8.2\pm0.4)\times10^{10}$\,K\,\kms\,pc$^{2}$, correcting from the
lensing magnification the intrinsic CO(3-2) luminosity is then $(3.7\pm
0.2)\times 10^{9}$\,K\,\kms\,pc$^{2}$.  The CO(3-2) luminosity we
measure is roughly half of the value found for Arp\,220 (Solomon
et al.\ 1997),  although Arp\,220 is comparably luminous
in the far-infrared to \smm1.

Using the relation from Solomon et al.\ (1997) and Downes \& Solomon
(1998) which indicate a CO luminosity to gas mass conversion factor of
$\sim$\,0.8--1.6\,M$_\odot$ (K\,{\kms}\,pc$^{2})^{-1}$ we derive a
molecular gas mass of $(4.5\pm 1.0)\times10^{9}$\,M$_\odot$, giving a
gas mass surface density of $\gs 10^3$M$_\odot$\,pc$^{-2}$.  Comparing
this to our dynamical estimates suggests a gas mass fraction of
$\sim60$\% for a disk/ring-like structure or $\sim30$\% for a
merger. Combined with our earlier estimate of the dust mass, this
indicates a dust-to-gas ratio of $0.004\pm 0.001$, similar to that seen
in local galaxies (Dunne et al.\ 2000).

The star-formation rate estimated from the far-infrared luminosity by
Kneib et al.\ (2004a) is $\sim$\,500\,M$_\odot$\,yr$^{-1}$ for stars
$>$\,0.1\,M$_\odot$ (note that in contrast the H$\alpha$-derived SFR
is only $\sim 10$\,M$_\odot$\,yr$^{-1}$, before correcting for
extinction and slit losses).  Taking the far-infrared estimate of the
SFR this indicates that the current activity cannot continue at its
present level for more than another 10\,Myrs.  Even allowing for the
fact that the far-infrared luminosity is dominated by the emission
from the most massive stars --  thus allowing us to use an IMF
dominated by massive stars or even to adopt a top-heavy IMF
(Baugh et al.\ 2005) -- would only extend this lifetime to
$\sim$\,50\,Myrs.  These estimates are comparable to the
crossing-time of the system, indicating that the starburst is likely
to be an instantaneous non-equilibrium event -- arguing against a
stable disk/ring configuration for the gas within this system -- and
suggesting instead that we are seeing the results of a tidal-induced
starburst in a pair of merging galaxies ($\alpha$ and $\beta$).  The
rest-frame UV/optical morphology of this system suggests that a
highly-obscured starburst is being triggered in the overlap region
between the two galaxies (component $\gamma$ in Kneib et al.\ 2004a)
as is commonly seen in low-redshift mergers.

We can also compare the ratio of the far-infrared luminosity produced
by \smm1\ to its H$_2$ gas mass: $L_{\rm FIR}/M_{\rm H_2} \sim 320$.
This is roughly ten time higher than seen in Arp\,220 (or comparably
luminous galaxies locally, Dunne et al.\ 2000; Downes et al.\ 1997).
This high ratio indicates that the activity in \smm1\ is far more
efficiently or vigorously using the available gas supply than is
typical in starburst systems at $z\sim 0$.  This may either result from
special physical conditions associated with intense starbursts at high
redshifts.  For example the activity could derive from wide spread
instabilities in large gas disks triggered by dynamical interactions,
which are not possible in comparably massive and gas-rich, but
bulge-stabilised, systems at low redshift.  Alternatively, the high
$L_{\rm FIR}/M_{\rm H_2}$ could reflect far-infrared emission from a
highly-obscured AGN -- although there are no spectral signatures of an
AGN.  Turning this argument around -- the free availability of gas to power
such nuclear activity, but the apparent lack of any signatures of an
AGN in the spectroscopy of this source from Kneib et al.\ (2004a), may
indicate that the central black hole(s) in this system (and the
associated bulge) has not yet grown large enough to produce a luminous
AGN.  The current interaction, and any subsequent mergers with the
other close companions in the compact and dynamically-cold group around
\smm1\ (the Lyman-break galaxies \#384 and \#273, Kneib et al.\ 2004a),
will provide the opportunity for substantial bulge growth and
associated accretion-driven activity from the central black hole of
this system.

To judge how the properties of this relatively low-luminosity SMG
compare to the other well-studied population of moderately active
galaxies at $z\gs 2$, the Lyman-break population (Steidel et al.\
2004), we can compare the parameters we measure for \smm1\ to those
measured for a similarly highly magnified Lyman-break galaxy: cB58 at
$z=2.73$ (Baker et al.\ 2001, 2004).  cB58 has a SFR of just
20\,M$_\odot$\,yr$^{-1}$, an inferred gas mass of $\sim 7\times
10^9$\,M$_\odot$, a dynamical mass of $\sim 1$--2$\times
10^{10}$\,M$_\odot$, a gas fraction of $\sim 0.3$--0.7 and an average
gas surface mass density of $\Sigma\sim 500$\,M$_\odot$\,pc$^{-2}$ (all
within a 2\,kpc radius, Baker et al.\ 2004).  These estimates are very
similar to those we derive for \smm1\ with the exception that \smm1\
has an order of magnitude higher instantaneous star-formation rate and
a marginally higher gas surface density.  Taken with our previous
suggestion that \smm1\ represent a merger, the comparison to cB58
suggests that it may be a merger between two typical Lyman-break
galaxies.  The merger has increased the gas density within the system
and triggered a burst of activity which will quickly use up the bulk of
the gas reservoir in a single event (which would have otherwise fueled
a more sedate level of activity for 100's Myrs).

In summary, several of the properties of \smm1\ appear to resemble
those of similar luminosity merging/interacting systems at low
redshift, e.g.\ the dust spectral index, dust-to-gas ratio, presence 
of an obscured starburst in the overlap zone between the merging components, 
etc.  However, the ratio of far-infrared luminosity to gas mass and the
physical extent of the apparently high-density gas reservoirs in
the progenitors are both an order of magnitude larger than for local
systems.  As \smm1\ is an order of magnitude fainter at submm wavelengths
than other distant SMGs studied in detail to date, and also less
massive, it is not yet clear how many of these similarities and
differences to low-redshift ULIRGs will carry over to the bulk of the
SMG population.  However, we stress that the properties of \smm1\ may
be more representative of the high-redshift population which produces a
large fraction of the submm cosmic extragalactic background in the
submm waveband, and hence are more relevant for understanding the
evolutionary behaviour of the majority of the oldest stellar components
in local galaxies: bulges.  The fact that the
properties of this system are more likely to resemble those expected for
a merger of two typical Lyman-break galaxies suggests a 
link between these two populations (at least for sub-mJy
SMGs). We also note that \smm1\ has yet to be
observed at the highest spatial resolution achievable with the IRAM
array -- these will provide a further factor of ten improvement in the
resolution -- probing the gas distribution within this rare example of
the sub-mJy submm galaxy population at sub-kpc scales,  which will
fully characterize the nature of this system.

\begin{acknowledgements}
We thank the anonymous referee for its comments which help to improve this
paper.  We acknowledge useful discussion with Rob Ivison and Colin Borys.
We also thank Michael Grewing for according us Director
Discretionary time which helped us to secure these results.  
JPK acknowledges support from CNRS and Caltech, and thanks IRAM staff
for its hospitality during his stay in Grenoble while working on these data. 
IRS acknowledges support from the Royal Society. A.W.B. acknowledges support 
from the NSF under grant AST-0205937, the
Research Corporation and the Alfred P. Sloan foundation.

\end{acknowledgements}

\end{document}